\documentstyle[psfig,onecolumn,referee]{mn2e}

\input{epsf}

\newif\ifAMStwofonts
\AMStwofontstrue

\newcommand{\go}{\mathrel{\raise.3ex\hbox{$>$}\mkern-14mu
             \lower0.6ex\hbox{$\sim$}}}
\newcommand{\lo}{\mathrel{\raise.3ex\hbox{$<$}\mkern-14mu
             \lower0.6ex\hbox{$\sim$}}}
\newcommand{\lp}{\left(}
\newcommand{\rp}{\right)}

\newcommand{\vecB}{\bmath B}
\newcommand{\vecmu}{\bmath \mu}
\newcommand{\vecW}{\bmath \Omega}
\newcommand{\vecE}{\bmath E}

\newcommand{\veck}{{\bmath k}}

\newcommand{\vechatk}{\hat{\bmath k}}
\newcommand{\vechaty}{\hat{\bmath y}}
\newcommand{\vechatz}{\hat{\bmath z}}

\newcommand{\vechatY}{\hat{\bmath Y}}
\newcommand{\vechatZ}{\hat{\bmath Z}}
\newcommand{\vechatB}{\hat{\bmath B}}
\newcommand{\vechatW}{\hat{\bmath \Omega}}

\newcommand{\singh}{\sin\theta_B}
\newcommand{\singhsq}{\sin^2\theta_B}
\newcommand{\cosgh}{\cos\theta_B}

\newcommand{\ggtheta}{\lp1-\beta\cosgh\rp}

\newcommand{\wc}{\omega_{\rm c}}

\newcommand{\polarb}{\beta_{\rm pol}}
\newcommand{\PA}{\phi_{\rm PA}}
\newcommand{\intd}{{\rm d}}

\newcommand{\phiFR}{\phi_{\rm FR}}
\newcommand{\RM}{{\rm RM}}

\title[]{Faraday Rotation in Pulsar Magnetosphere}
\author[C. Wang, J.L. Han and D. Lai]
  {Chen Wang$^{1}$, J. L. Han$^{1}$, Dong Lai$^{2}$ \\
  $^{1}$ National Astronomical Observatories, Chinese Academy of
  Sciences.  A20 Datun Road, Chaoyang District, Beijing 100012, China \\
  $^{2}$ Department of Astronomy, Cornell University, Ithaca, NY
  14853, USA \\
  {\rm E-mail: wangchen@nao.cas.cn}}
\begin{document}

\maketitle

\label{firstpage}

\begin{abstract}

 The magnetosphere of a pulsar is composed of relativistic plasmas
 streaming along the magnetic field lines and corotating with the
 pulsar. We study the intrinsic Faraday rotation in the pulsar
 magnetosphere by critically examining the wave modes and the
 variations of polarization properties for the circularly polarized
 natural modes under various assumptions about the magnetosphere
 plasma properties.  Since it is difficult to describe analytically
 the Faraday rotation effect in such a plasma, we use numerical
 integrations to study the wave propagation effects in the corotating
 magnetosphere. Faraday rotation effect is identified among other
 propagation effects, such as wave mode coupling and the cyclotron
 absorption. In a highly symmetrical electron-positron pair plasma,
 the Faraday rotation effect is found to be negligible. Only for
 asymmetrical plasmas, such as the electron-ion streaming plasma, can
 the Faraday rotation effect become significant, and the Faraday
 rotation angle is found to be approximately proportional to
 $\lambda^{0.5}$ instead of the usual $\lambda^2$-law. For such
 electrons-ion plasma of pulsar magnetosphere, the induced rotation
 measure becomes larger at higher frequencies, and should have
 opposite signs for the emissions from opposite magnetic poles.

\end{abstract}

\begin{keywords}
polarization -- radiative transfer -- star: magnetic fields -- pulsars:
general
\end{keywords}

\section{Introduction} \label{sec:intro}

Radio emissions from pulsars are generally highly polarized. When the
radio waves propagate through the magneto-ionized interstellar medium
(ISM), the linear polarization plane is rotated by an angle
proportional to $\int Bn_e\intd l\cdot\lambda^2$. This is the Faraday
rotation (FR) effect. The rotation measure, $\RM=\int B n_e \intd l$,
can be measured from the polarization angles at different frequencies
(Hamilton \& Lyne 1987; Han et al. 1999, 2006).

Some questions need to be answered when pulsar rotation measures are
used to investigate the properties of the interstellar medium.  Do the
$\RM$s of pulsars completely come from the ISM? Is there any intrinsic
Faraday rotation from pulsar magnetosphere? If so, how large is it?
Because the intrinsic polarization position angles of pulsar emission
vary with the pulsar rotation phase, is the RM different for different
phase bin?  How to average the RMs by different weights of different
phases?  The RMs from ISM should not vary with rotation
phase. However, recent observations by Ramachandran et al. (2004) and
Noutsos et al. (2009) showed that for some pulsars the observed RMs do
vary with the rotation phases, indicating that the RMs do not just
come from the ISM.

There are three possible reasons for the different phase-resolved RM:
(1) the effect of incoherent superposition of quasi-orthogonal
polarization modes of pulsar emission (Ramachandran et al. 2004);
(2) interstellar scattering (Noutsos et al. 2009, Karastergious 2009);
(3) intrinsic Faraday rotation in pulsar magnetospheres. The first two
possibilities have already been discussed previously, while the third
has not been addressed carefully. Although Kennett \& Melrose (1998)
discussed the generalized Faraday rotation effect in relativistic
plasmas when the natural modes are linearly polarized, they did not
consider the Faraday rotation for the circularly polarized natural
modes in such a plasma.

In this paper, we investigate the intrinsic Faraday rotation in pulsar
magnetospheres when the natural wave modes are circularly polarized. A
pulsar magnetosphere consists of highly magnetized relativistic
streaming plasmas, different from the ISM. Thus, the simple $\lambda^2$
relation for the Faraday rotation may not apply. In \S~2, we give the
theoretical description of the natural modes in the relativistic
streaming plasma, and analyze the Faraday rotation effect. Because it
is not easy to obtain analytic expression for the Faraday rotation effect
for such a plasma, in \S~3 we study this effect by numerically integration
of the wave vectors in a self-consistent manner. Our conclusions
are presented in \S~4.

\section{Faraday rotation for circularly polarized natural modes in
Pulsar Magnetosphere }

In any magnetized medium, when the natural modes are circularly
polarized and the mode evolution is adiabatic, Faraday rotation
effect should rotate the linear polarization position angle
($\PA$) by
\begin{equation}
\Delta \PA= \int\frac{\Delta k}{2}\intd r. \label{eq:dPA}
\end{equation}
Here $\Delta k=\Delta n\omega/c$ is the wave number difference of the
two circularly polarized natural modes, $\intd r$ is the unit distance
in photon ray \footnote{In this paper, $r$ stands for the distance
  from the neutron star center. Since the emission and propagation
  regions are far away from the NS surface, we do not distinguish the small difference of
  $\intd r$ between the photon ray and radial direction.}. In our
previous papers (Wang \& Lai 2007; Wang, Lai \& Han 2010,
hereafter WLH10), we have already derived the dielectric tensor
and wave modes in pulsar magnetospheres composed of streaming,
relativistic plasmas of various compositions and Lorentz factors.
With the dielectric tensor, we can solve the Maxwell equation and
obtain two eigenmodes, to be labeled as the plus ``+'' mode and
minus ``$-$'' mode, and corresponding refractive indices $n_\pm$.
We write the polarization vectors of the two modes as
$\vecE_{\pm}=\vecE_{\pm T}+\vecE_{\pm z}\vechatz$ in the
$xyz$-frame, with $ \vechatz = \vechatk$, $\vecB$ in the
$xz$-plane and $\vechatk \times \vechatB = - \sin\theta_B
\vechaty$. The transverse part of the mode polarization vector is
given by
\begin{equation}
\vecE_{\pm T}=\frac{1}{(1+K_{\pm}^2)^{1/2}}(iK_{\pm},1),
\label{eq:eigenvector}
\end{equation}
where $iK_\pm=(E_x/E_y)_\pm$. From the Maxwell equation and dielectric
tensor, we obtain
\begin{equation}
K_\pm = \polarb\pm\sqrt{\polarb^2+1},
\label{eq:K}
\end{equation}
with the polarization parameter
\begin{equation}
\polarb \simeq -\frac{f_\eta\singhsq u_r^2/(1-u_r^2)}{2\Sigma_s
f_{12,s}(\cosgh-\xi_s\singh)}. \label{eq:polarb1}
\end{equation}
Here $f_\eta$, $f_{12,s}$, $\xi_s$ are factors in dielectric
tensor [see eqs.~(3.15) and (3.16) of WLH10], the sum $\Sigma_s$
runs over each charged particle species ``s'' (electrons,
positrons and/or ions). The dimensionless parameter
$u_r=\wc/[\gamma\omega\ggtheta]$ stands for the ratio of cyclotron
frequency $\wc$ and the rest-frame frequency
$\gamma\omega\ggtheta$, with $\omega$ the wave frequency, $\gamma$
the Lorentz factor of streaming plasma, and $\theta_B$ the
$\veck$-$\vecB$ angle. Obviously, when $|\polarb|\gg 1$, the two
eigenmodes are linearly polarized, while for $|\polarb| \ll 1 $
the two modes are circularly polarized ($K_\pm=\pm1$).  The
refractive indices of the two modes can be written as
\begin{equation}
n_\pm^2=1+f_{11}\pm K_\pm\Sigma_s f_{12,s}(\cosgh-\xi_s\singh).
\label{eq:n1}
\end{equation}
Since $n_\pm$ is very close to the unity, the difference between
the two refractive indices can be written as
\begin{equation}
\Delta n\simeq \sqrt{1+\polarb^2}\Sigma_s f_{12,s}(\cosgh-\xi_s\singh).
\label{eq:n1}
\end{equation}

If the electrons and positrons in the plasma are purely symmetrical in
density and velocity, then the denominator of eq.~(\ref{eq:polarb1})
equals zero and $\polarb=\infty$ along the photon ray; this means that
the two natural modes are always linearly polarized. However, in
reality there always exists some asymmetry between electrons and
positrons in the pulsar magnetosphere. For example, in the
Goldreich-Julian magnetosphere model (Goldreich \& Julian 1969), there
exists a net charge density, called Goldreich-Julian density, $N_{\rm
  GJ}$, which specifies the density difference between electrons and
positrons.  Ions (such as H$^+$ or Fe ions) may also exist in the
magnetosphere. Since the interaction between ions and photons is much
weaker than that between electrons/positons and photons, the existence
of ions enhances the asymmetry of the plasma. The Lorentz factors of
electrons and positrons may also be different because of different
electric field acceleration (Arons 1983; Kazbegi et al. 1991). If the
asymmetry between electrons and positrons is sufficiently large, even
though the natural modes in the inner magnetosphere of pulsars are
linearly polarized, when the photons propagate to the outer
magnetosphere, the natural modes could become elliptically or
circularly polarized.

In this paper, for simplicity we consider a cold pair plasma, in
which electrons and positrons have the same Lorentz factors
($\gamma_p=\gamma_e=\gamma$) but different densities, $\Delta
N=N_p-N_e\neq 0$.  The density difference could be
Goldreich-Julian density, but we will not entirely restrict
ourselves to this constraint. In this case all the equations are
simplified. The polarization parameter $\polarb$ becomes
\begin{equation}
\polarb = -\frac{u_r\singhsq\gamma^{-2}}{2(1-\beta\cosgh)(\cosgh-\beta)\Delta N/N},
\label{eq:polarb2}
\end{equation}
with the plasma velocity $\beta=1/\sqrt{1+\gamma^2}$ and the total
plasma density $N=N_p+N_e$. The difference of refractive indices
can also be simplified to
\begin{equation}
\Delta n\simeq -\sqrt{1+\polarb^2}\frac{v\gamma^{-1}u_r}{1+i2\gamma_{\rm rad}-u_r^2}
    \frac{\Delta N}{N}\frac{\cosgh-\beta}{1-\beta\cosgh}.
\label{eq:n2}
\end{equation}
Here $v=\omega_{\rm p}^2/\omega^2$ relates to the plasma frequency
$\omega_{\rm p}$, and the radiative damping parameter
\begin{equation}
\gamma_{\rm rad}=\frac{4e^2\omega_{\rm c}}{3m_ec^2}
\end{equation}
is important only near the cyclotron resonance and can be
neglected away from the resonance. The cyclotron resonance radius
$r_{\rm cyc}$ can be written as [see
  eq.~(4.49) of WLH10]
\begin{equation}
r_{\rm cyc}/R_\ast= 1.8\times10^3B_{\ast12}^{1/3}\nu_9^{-1/3}\theta_B^{-2/3}
\gamma^{-1/3}. \label{eq:r_cyc}
\end{equation}
Here the magnetic field is described by $B(r)\simeq
B_\ast(R_\ast/r)^3$, $R_\ast$ is the radius of the neutron star
(usually we set $R_\ast=10$\,km), $B_\ast=10^{12}B_{\star 12}$\,G is
the surface magnetic field, $\nu=\nu_9$\,GHz is the wave frequency,
and $r$ is the distance from the neutron star center.  Typically,
$\theta_B\ll 1$, $\gamma \gg1$ and $\theta_B\gamma\gg1$,
equations~(\ref{eq:polarb2}) and (\ref{eq:n2}) can be simplified
further:
\begin{equation}
\polarb\simeq \frac{2u_r}{\theta_B^2\gamma^2\Delta N/N},
\label{eq:polarb3}
\end{equation}
\begin{equation}
\Delta n \simeq \sqrt{1+\polarb^2}\frac{v\gamma^{-1}u_r}{1-u_r^2}
    \frac{\Delta N}{N},
\label{eq:n3}
\end{equation}
where we have neglected the radiative damping term. If
$\theta_B^2\gamma^2\Delta N/N\gg1$, the natural modes could become
circularly polarized even before the cyclotron resonance (where
$u_r=1$),  and Faraday rotation could affect the photon linear
polarization angle. Before and after the cyclotron resonance
(where $u_r=1$), $\Delta n$ will change signs, while $\polarb$
keeps the same sign. This means that the Faraday rotations to PA
have opposite directions before and after cyclotron resonance.

\subsection{Faraday rotation effect for circularly polarized natural modes}
 For a typical $\theta_B$ (not so close to 0, $\theta_B\gamma\gg1$),
 only when $u_r<\theta_B^2\gamma^2\Delta N/N$, the eigenmodes become
 circularly polarized [see eq.~(\ref{eq:polarb3})]. We define the
 radius of circularization $r_{\rm cir}$ by $|\polarb(r=r_{\rm cir})|
 = 1$, so that the normal modes become circularly polarized
 when $r> r_{\rm cir}$. According to eq.~(\ref{eq:polarb3}), the
 radius of circularization is
\begin{equation}
r_{\rm cir}/R_\ast= 2.2\times10^3B_{\ast12}^{1/3}P_{1s}^{-1}\nu_9^{-1/3}\theta_B^{-4/3}
\gamma^{-1}(\Delta N/N)^{-1/3}. \label{eq:r_cir}
\end{equation}
For the typical parameters of a pulsar magnetosphere,
$B_\ast=10^{12}$G, $P$=1s, $\nu$=1GHz, $\theta_B\sim 0.1$,
$\gamma=100$, we have $r_{\rm cir}/r_{\rm LC}\sim 0.1$. If the
mode evolution is adiabatic and refractive indices difference
$\Delta n$ is large enough, the linear polarization position angle
will be affected by the Faraday rotation significantly. The
position angle is rotated by
\begin{equation}
\Delta \PA\simeq \int_{r_{\rm cir}}^{r_{\rm LC}}\frac{\Delta k}{2}\intd r
= \int_{r_{\rm cir}}^{r_{\rm LC}}\frac{\Delta n\omega}{2c}\intd r
\simeq \int_{r_{\rm cir}}^{r_{\rm LC}} -\frac{v\gamma^{-1}u_r}{1-u_r^2}
    \frac{\Delta N}{N}\frac{\omega}{2c} \intd r, \label{eq:dPA}
\end{equation}
where we have assumed $\polarb\sim0$ since the natural modes are
circularly polarized. This integral cannot be evaluated
analytically. The most important variable is $B$, since $B\propto
r^{-3}$, and $v\propto B$, $u_r\propto B$. The other parameters in
the equation do not change or change slowly with $r$. Thus,
approximately we get
\begin{equation}
\Delta \PA\simeq
-\frac{v\gamma^{-1}}{u_r}\frac{\Delta N}{N}\frac{\omega}{2c} r_{\rm cyc}
\int_{r_{\rm cir}/r_{\rm cyc}}^{r_{\rm LC}/r_{\rm cyc}} \frac{u_r^2}{1-u_r^2}
     \intd \left(\frac{r}{r_{\rm cyc}}\right). \label{eq:dPA2}
\end{equation}
Suppose $x=r/r_{\rm cyc}$, then $u_r\simeq x^{-3}$, and the integration
term of eq.~(\ref{eq:dPA2}) becomes
\begin{equation}
F_x=\int_{r_{\rm cir}/r_{\rm cyc}}^{r_{\rm LC}/r_{\rm cyc}}
\frac{x^{-6}}{1-x^{-6}}\intd x. \label{eq:Fx}
\end{equation}
As discussed above, $\Delta n$ changes signs when photon
propagates across $r_{\rm cyc}$ (where $x=1$).
Then eq.~(\ref{eq:dPA2}) becomes
\begin{equation}
\Delta \PA\simeq 0.19\eta P_{\rm
1s}^{-1}B_{\ast,12}^{1/3}\theta_B^{4/3} \gamma^{-1/3}(\Delta
N/N)F_x\nu_9^{-1/3}, \label{eq:dPA3}
\end{equation}
where $\eta=N/N_{\rm GJ}$ is the plasma density parameter, $P_{\rm
1s}=P/1\,$s. Equation~(\ref{eq:dPA3}) gives a simple approximate
description to the final Faraday rotation angle. It is
proportional to $B_\ast$, $\eta$, $\Delta N/N$, but inversely
proportional to $P$, $\gamma$ and $\nu$. Notice that $\Delta
\PA\propto \nu_9^{-1/3}\propto\lambda^{1/3}$, quite different from
the FR in ISM (for which $\Delta \PA\propto\lambda^{2}$). If the
plasma is highly symmetrical ($\Delta N/N\ll 1$),
since $\PA\propto \Delta N/N$, the Faraday rotation is negligible.
However, if the plasma is highly asymmetrical, for example, $\Delta N/N\sim 1$
(the case of pure electrons), $r_{\rm cir}/r_{\rm cyc}$ could be less
than 1, the Faraday rotation angle could be significant. For
typical parameters of a pulsar magnetosphere, $B_\ast=10^{12}$ G,
$P=1$ s, $\gamma=100$, $\eta=1000$, $\Delta N/N\sim 1$,
$\theta_B\sim 0.1$, the circularization radius can be $r_{\rm
cir}/r_{\rm cyc}\sim 0.5$, and the final rotation angle $\Delta
\PA\simeq 0.9\nu_9^{-1/3}$.

Notice that in our derivation of equations~(\ref{eq:dPA2}) and
(\ref{eq:dPA3}), the $\veck$-$\vecB$ angle $\theta_B$ is assumed
to be a constant in the propagation.  In reality, $\theta_B$ also
changes along the photon path. Additionally, $F_x$ given by
eq.~(\ref{eq:Fx}) is related to $r_{\rm cir}$ and $r_{\rm cyc}$,
both of which are determined by the plasma parameters [see
  eqs.~(\ref{eq:r_cir}) and (\ref{eq:r_cyc})].  Thus,
the Faraday rotation angle given in equations~(\ref{eq:dPA2}) and
(\ref{eq:dPA3}) is only a simple approximation. We will obtain the
precise rotation angle by numerical integration of wave evolution
equation in \S~3.

\subsection{Evolution of mode amplitude and adiabatic condition}

In the $xyz$ frame [defined with $\vechatz = \vechatk$, $\vechatB
= (-\sin\theta_B, 0, \cos\theta_B)$], there are two wave modes:
``+'' mode and ``$-$'' mode. We introduce a mixing angle,
$\theta_m$, via $\tan\theta_m=1/(K_+)$, so that
\begin{equation}
\tan 2\theta_m = \polarb^{-1}.\label{eq:theta_m}
\end{equation}
In the $xyz$ frame, the transverse components of the mode
eigenvectors are then
\begin{equation}
\vecE_+=\left(\begin{array}{c}i\cos\theta_m\\\sin\theta_m\end{array}\right),
\quad
\vecE_-=\left(\begin{array}{c}-i\sin\theta_m\\\cos\theta_m\end{array}\right).
\end{equation}
In the fixed observers' $XYZ$ frame [defined by $\vechatZ =
\vechatk$, $\vecW$ in the $XZ$-plane and $\vechatk\times\vechatW =
\Omega\sin\zeta\vechatY$, the direction of $\vecB$ in this frame
is ($\theta_B$, $\phi_B$)], they become
\begin{equation}
\vecE_+=\left(\begin{array}{c}i\cos\theta_m\cos\phi_B-\sin\theta_m\sin\phi_B\\
                              i\cos\theta_m\sin\phi_B+\sin\theta_m\cos\phi_B\end{array}\right),\\
\vecE_-=\left(\begin{array}{c}-i\sin\theta_m\cos\phi_B-\cos\theta_m\sin\phi_B\\
                              -i\sin\theta_m\sin\phi_B+\cos\theta_m\cos\phi_B\end{array}\right).
\end{equation}
The general wave amplitudes can be written as
\begin{equation}
\left(\begin{array}{c}A_{X}\\A_{Y}\end{array}\right)
=A_{+}\vecE_++
 A_{-}\vecE_-.\\
\end{equation}
Substitute this into the wave equation, we obtain the mode amplitude
evolution equation:
\begin{equation}
i{\intd\over \intd r}\left(\begin{array}{c}A_+\\A_-\end{array}\right)
=\left[\begin{array}{cc}
-\Delta k/2+\phi_B'\sin2\theta_m & i\theta_m'+\phi_B'\cos2\theta_m \\
-i\theta_m'+\phi_B'\cos2\theta_m & \Delta k/2-\phi_B'\sin2\theta_m
\end{array} \right]
\left(\begin{array}{c}A_+\\A_-\end{array}\right),
\label{eq:me}
\end{equation}
where the superscript ($'$) specifies $\intd/\intd r$, $\Delta k=
k_+-k_-=\Delta n\omega/c$, and a non-essential unity matrix has
been subtracted. This equation generalizes the results in special
cases (where only $\theta_m$ or $\phi_B$ varies) studied in Lai \&
Ho (2002, 2003) and van Adelsberg \& Lai (2006), and therefore is
useful for understanding the effect of mode coupling and Faraday
rotation.

When the natural modes are circularly polarized, which means
$\theta_m=45^o$, eq.~(\ref{eq:me}) can be simplified to
\begin{equation}
i{\intd\over \intd r}\left(\begin{array}{c}A_+\\A_-\end{array}\right)
=\left[\begin{array}{cc}
-\Delta k/2+\phi_B' & i\theta_m' \\
-i\theta_m' & \Delta k/2-\phi_B'
\end{array} \right]
\left(\begin{array}{c}A_+\\A_-\end{array}\right).
\label{eq:me_cp}
\end{equation}
The adiabatic parameter is then defined as
\begin{equation}
\Gamma_{\rm ad}=\left| \frac{\Delta k/2-\phi_B'}{\theta_m'}\right|.
\end{equation}
Only when $\Gamma_{\rm ad}\gg1$, the mode evolution is adiabatic,
so that the two natural modes can propagate separately and the
Faraday rotation can occur. For the non-adiabatic case of
$\Gamma_{\rm ad}\ll1$, the two modes are coupled with each other,
the electromagnetic waves do not interact with the medium and the
wave polarization states keep unchange.

For the wave propagation in the interstellar medium, the adiabatic
condition is very easily satisfied. However, the pulsar
magnetosphere is filled with relativistic streaming pair plasma,
the condition is only satisfied in some regions. To evaluate the
adiabatic condition $\Gamma_{\rm ad}\gg 1$ for $r\go r_{\rm cir}$
(or $\polarb\sim0$), we use $\Delta k=\Delta n \omega /c$, with
$\Delta n$ given by eq.~(\ref{eq:n3}), and
\begin{equation}
\phi_B'=F_\phi/r_{\rm LC}=2.1\times10^{-10}F_\phi P_{1s}^{-1}~{\rm cm}^{-1},
\end{equation}
with
\begin{equation}
F_{\phi} =
\frac{\sin^2\alpha\cos\zeta-\sin\alpha\cos\alpha\sin\zeta\cos\Psi}
{1-\left(\cos\alpha\cos\zeta+\sin\zeta\sin\alpha\cos\Psi\right)^2}.
\end{equation}
Here $\alpha$ is the inclination angle between the magnetic axis
and the rotation axis of a pulsar, $\zeta$ is the view angle from
the rotation axis, and $\Psi=\Psi_{\rm i}+\Omega t$ is the pulsar
rotation phase. From the definition of $\theta_m$, we have
\begin{eqnarray}
\theta_m'&=&-\frac{1}{2}\sin^22\theta_m\polarb\frac{\polarb'}{\polarb}\nonumber\\
&=&1.1\times10^{-11}B_{\ast12}P_{1s}^{-4}\nu_9^{-1}\left(\frac{r}{r_{\rm LC}}\right)^{-3}
  \left(\frac{L_\beta}{r_{\rm LC}}\right)^{-1}\frac{\sin^22\theta_m}{\theta_B^{4}\gamma^{3}(\Delta N/N-\Delta\gamma/\gamma)}.
\end{eqnarray}
Here we define $L_\beta=\polarb/\polarb'$ which is the scale length of
$\polarb$ variation; generally $L_\beta\sim r$ since $\polarb \propto
r^{-6}\theta_B^{-2}$. In the upper panels of Figure \ref{fig:single},
we give some examples for the evolution of the quantities $\Delta
k/2$, $\phi_B'$ and $\theta_m'$ along the photon ray for different
plasma parameters. In the highly symmetrical plasma model (see
Fig.~\ref{fig:single}a), the refractive indices difference $\Delta n$
is small, and the evolution is non-adiabatic ($\Delta k/2\lo
\phi_B',~\theta_m'$).  Thus there is no Faraday rotation in this
model. On the other hand, for the highly asymmetrical plasma model
(see Fig.~\ref{fig:single}b), the evolution remains adiabatic all the
way along the photon path ($\Delta k/2\gg \phi_B'$, $\theta_m'$).
Therefore we expect that in this (asymmetric) plasma model, Faraday
rotation will be generated. Only in some special cases, for example,
if a photon propagates through the region where the magnetic fields
change signs along the path very quickly, $\phi_B'$ could be larger
than $\Delta k/2$, so that the non-adiabatic mode evolution could
occur. This ``quasi-transverse magnetic field'' case was considered by
Broderick \& Blandford (2009) and Melrose (2010). However, the radio
emission we considered in this paper is generated near the tangential
direction of the magnetic field line, and magnetic fields in the outer
regions do not change sign along the photon path, thus the
``quasi-transverse'' situation does not occur in pulsar magnetosphere.

\section{Numerical Results}

The analytic description above for the physical process of the
Faraday rotation can only give qualitative results under
simplified conditions. Here we use the numerical integrations to
calculate how radio polarization evolves as the wave propagates
through the pulsar magnetosphere. This allows us to obtain the
exact values for the rotated polarization angles $\PA$ and the
rotation measure RM.

\subsection{Single ray evolution}

It is generally accepted that pulsar radio emission originates
from the open field line region at a few to tens of neutron star
radii (e.g. Cordes 1978; Blaskiewicz et al. 1991; Kramer et
al.~1997; Kijak \& Gil 2003). We assume the emission height
$r_{\rm em}=50R_\ast$ and at the emission point, the photon is
polarized in the $\veck$-$\vecB$ plane (i.e. the O-mode in the
case of curvature radiation) and propagates along the tangential
direction of the local field line. We neglect the emission cone of
opening angle $1/\gamma$. For a given emission height $r_{\rm
em}$, pulsar initial rotation phase $\Psi_{\rm i}$, the
inclination angle $\alpha$ (i.e., the $\vecmu$-$\vecW$ angle), the
view angle of line of sight $\zeta$ (i.e., the $\veck$-$\vecW$
angle), the surface magnetic field $B_\ast$, and the plasma
properties (plasma density parameter $\eta=N/N_{\rm GJ}$, Lorentz
factor of the streaming plasma $\gamma$), we can numerically
calculate the dielectric tensor at each point along the photon ray
[see equation~(3.15) in WLH10], and integrate the wave evolution
equation [given by equation~(2.10) in WLH10] from the emission
point to a very large radius, which we take the light cylinder
radius $r_{\rm lc}$, to determine the final polarization state of
the photon.

\begin{figure}
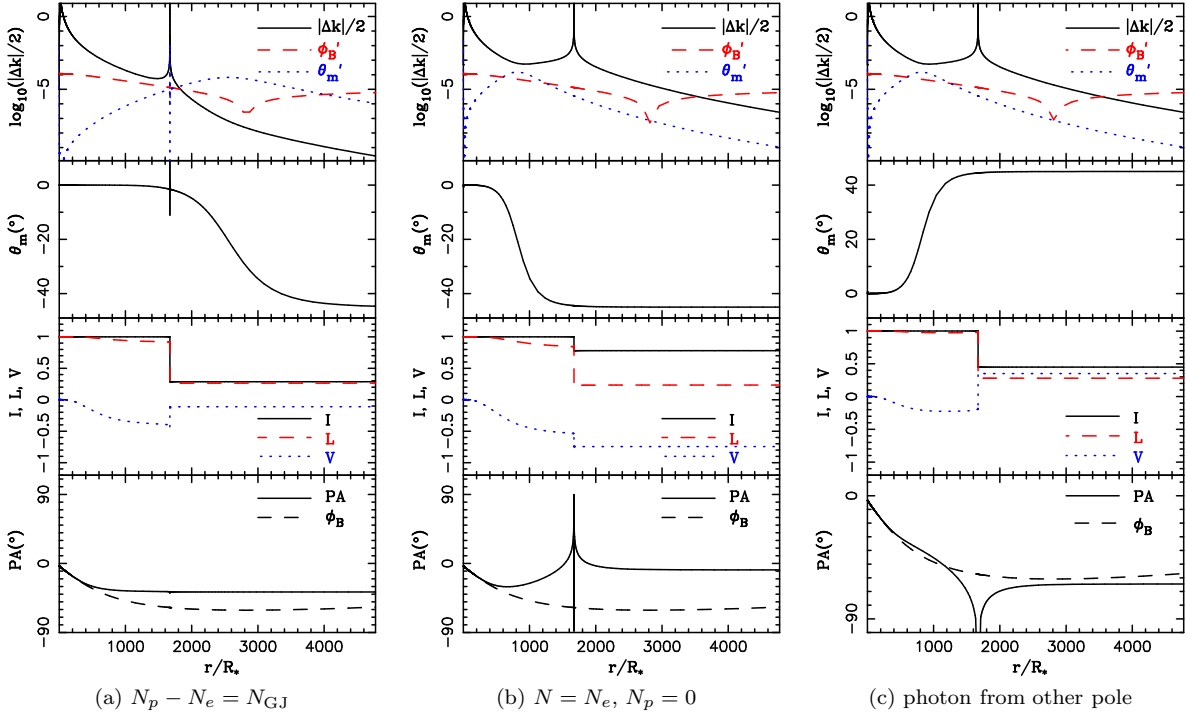

\begin{tabular}{ccc}
\psfig{figure=f1a.ps,angle=-90,height=9cm} &
\psfig{figure=f1b.ps,angle=-90,height=9cm} &
\psfig{figure=f1c.ps,angle=-90,height=9cm} \\
(a) $N_p-N_e=N_{\rm GJ}$ & (b) $N=N_e$, $N_p=0$ &
(c) photon from other pole \\
\end{tabular}
\caption{Single photon evolutions across the pulsar magnetosphere
with the plasma density parameter $\eta=N/N_{\rm GJ}=1000$ and the
Lorentz factor $\gamma=100$ for a pulsar with surface magnetic
field $B_\ast=10^{12}$\,G and spin period $P=1$\,s. The magnetic
inclination angle is $\alpha=30^o$ and the impact angle is
$\chi=\zeta-\alpha=5^o$. The calculations were made for the wave
frequency $\nu=1$\,GHz, initial rotation phase $\Psi_{\rm i}=0^o$,
and emission height $r_{\rm em}=50R_\ast$. Panel (a) is for the
pair plasma with $N_p-N_e=N_{\rm GJ}$. Panel (b) is the case for
the magnetosphere plasma consisting of electrons and ions (without
positrons), $N=N_e$, $N_p=0$. Panel (c) is almost the same as
panel (b) but for the emission of the opposite magnetic pole.}
\label{fig:single}
\end{figure}

We first consider magnetospheres consisting of electrons and
positrons, with the net charge density given by the
Goldreich-Julian model, i.e., $N_p-N_e=N_{\rm GJ}$. We assume that
the total plasma density is $N=N_p+N_e=1000N_{\rm GJ}$, which
means the plasma is almost symmetric for electrons and positrons.
Figure~\ref{fig:single}a shows an example of the photon
polarization evolution along its trajectory in such a plasma. The
wave natural modes are linearly polarized within the cyclotron
absorption radius ($\sim1650R_\ast$). Before the cyclotron
resonance, the dominated propagation effect is the wave mode
coupling (around $\sim600R_\ast$), where circular polarization is
generated. Although the natural modes become circularly polarized
near $r\simeq 2600R_\ast$, the refractive index difference is very
small and the mode evolution is non-adiabatic too ($\Delta k/2 <
\phi_B',~\theta_m'$); these make the Faraday rotation ineffective.
Thus the PA remains unchanged outside the wave mode coupling area
of $\sim600R_\ast$ (see the bottom panel of
Fig.~\ref{fig:single}a).

Figure~\ref{fig:single}b shows the case where the magnetosphere
plasma consists of electrons and ions, without positrons at all,
$N=N_e=1000N_{\rm GJ}$, $N_p=0$. The ions can be neglected for the
emission and propagation effects since they are too heavy compared
with the electrons. In this case, the adiabatic condition of
$\Delta k/2 > \phi_B',~\theta_m'$ is always satisfied (the upper
panel of Fig.~\ref{fig:single}b). The refractive index difference
is also much larger than that for the symmetric case. In the
region of $r<800R_\ast$, the wave mode coupling dominates for the
wave propagation. Farther out, the natural wave modes becomes
circularly polarized ($r_{\rm cir}\sim 900R_\ast$), Faraday
rotation occurs and the direction of linear polarization varies
(the bottom panel). Note that at the cyclotron resonance ($r_{\rm
cyc}\sim 1700 R_\ast$), $\Delta k/2$ changes sign, so the rotation
measure changes sign too. The final Faraday rotation angle is
about $20^o$. In Fig.~\ref{fig:single}c we consider the same case
as in Fig.~\ref{fig:single}b except for the emission coming from
the other magnetic pole.

\subsection{Rotation measure in electron-ion plasma of pulsar magnetosphere}

As discussed above, the Faraday rotation effect is negligible in a
highly symmetric electron-positron pair plasma of pulsar
magnetosphere, while may be sigificant in an asymmetric plasma,
e.g., electron-ion streams.  In this subsection, we study the
asymmetric case in more detail and present the rotation measure values
for various plasma parameters.

\begin{figure}
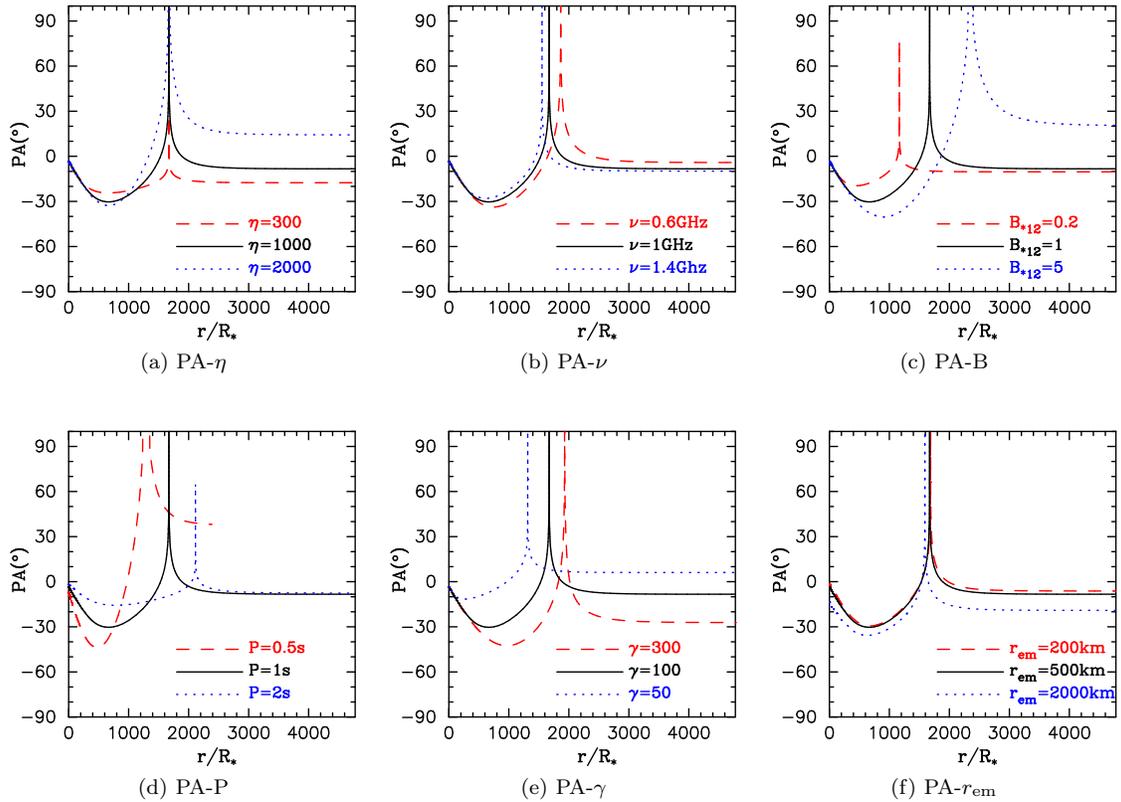

\centering
\begin{tabular}{ccc}
\psfig{figure=single_eta.ps,angle=-90,height=4.5cm} &
\psfig{figure=single_nu.ps,angle=-90,height=4.5cm} &
\psfig{figure=single_B.ps,angle=-90,height=4.5cm} \\
(a) PA-$\eta$ & (b) PA-$\nu$ & (c) PA-B \\
&&\\
&&\\
\psfig{figure=single_P.ps,angle=-90,height=4.5cm} &
\psfig{figure=single_gamma.ps,angle=-90,height=4.5cm} &
\psfig{figure=single_EH.ps,angle=-90,height=4.5cm} \\
 (d) PA-P & (e) PA-$\gamma$ & (f) PA-$r_{\rm em}$
\end{tabular}
\caption{The evolution of PA along the photon ray for various
pulsar and magnetosphere plasma properties, including density
$\eta=N/N_{\rm GJ}$ (panel a), frequency (panel b), surface B
field (panel c), rotation period (panel d), Lorentz factor (panel
e) and initial emission height (panel f). In each panel, only one
parameter is varied and labeled, while other parameters are the
same as those used in Fig.~\ref{fig:single}b.}
\label{fig:single_multi}
\end{figure}

In Figure~\ref{fig:single_multi}, we show some examples of the PA
changes for different pulsar and magnetosphere plasma properties,
such as different electron density parameter $\eta=N/N_{\rm GJ}$
(Fig.~\ref{fig:single_multi}a), wave frequency $\nu$
(Fig.~\ref{fig:single_multi}b), surface magnetic field $B_\ast$
(Fig.~\ref{fig:single_multi}c), pulsar period $P$
(Fig.~\ref{fig:single_multi}d), Lorentz factor of plasmas $\gamma$
(Fig.~\ref{fig:single_multi}e) and the emission height $r_{\rm em}$
(Fig.~\ref{fig:single_multi}f). In general, the Faraday rotated PA
increases with $\eta$ and $B_\ast$, but decreases with $\nu$ and $P$,
and almost does not change with $\gamma$ and $r_{\rm em}$, roughly in
agreement with eq.~(\ref{eq:dPA3}).

Interpulses have been detected from a few tens of pulsars (e.g.
Weltevrede \& Johnston 2008), which is 180$^o$ away from the main
pulse and comes from the other magnetic pole of a pulsar. In this
case, the magnetic fields along the photon trajectory have the
opposite directions, and Faraday rotation should have the opposite
sign. The PA evolution is also exactly opposite to that for the
mainpulse case (see Fig.~\ref{fig:single}c).

\subsubsection{Faraday Rotation Measure}

\begin{figure}
\centerline{\psfig{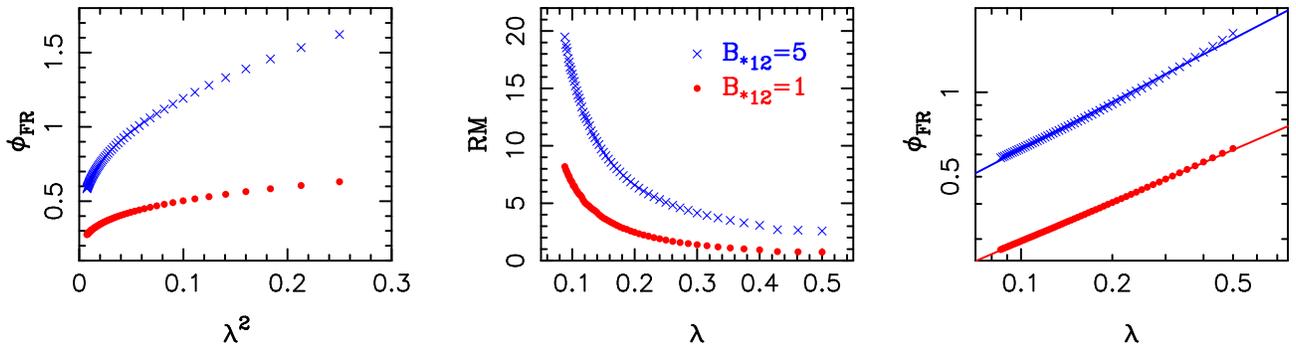}}
\caption{The Faraday rotated angle produced in the magnetosphere
as a function of the wavelength for $B_{\ast}=10^{12}$G
(``$\bullet$'') and $B_{\ast}=5\times10^{12}$G (``$\times$''). The
left panel shows $\phiFR$ against $\lambda^2$ and the middle panel
shows RM (=$\intd\phiFR/\intd\lambda^2$) against $\lambda$. The
right panel shows the possible relation between log$_{10}(\phiFR)$
and log$_{10}$($\lambda$), and the two solid lines are
$\phiFR=0.86\,\lambda^{0.47}$ for $B_\ast=10^{12}$G and
$\phiFR=2.28\lambda^{0.56}$ for $B_\ast=5\times10^{12}$G. The
pulsar and plasma parameters are the same as used in
Figure~\ref{fig:single}b, except for the varying frequencies.}
\label{fig:PAvsFreq}
\end{figure}

For given pulsar and plasma properties, we can calculate the
rotated PA for various frequencies.  Note that some other
propagation effects, such as wave mode coupling and
quasi-tangential effect (see WLH10), can also modify PA unless the
modes evolution becomes non-adiabatic or the natural modes become
circularly polarized. In general, the final PA value compared with
the initial value at emission point is given by
\begin{equation}
\PA = \phi_{\rm other}+\phiFR,
\end{equation}
where $\phiFR$ is the Faraday rotation angle, and $\phi_{\rm other}$ is
the rotation angle caused by other propagation effects in the
magnetosphere. The Faraday rotation angles have the same magnitude but
different signs for the opposite magnetic poles. The other propagation
effects, such as wave mode coupling, are the same for the opposite
poles. Note that the cyclotron absorption effect will absorb opposite
circular polarization but does not affect the linear polarization
angle. Therefore we have $\phi_{\rm PA,opp}=\phi_{\rm other}-\phiFR$
for the emission from the opposite magnetic pole. In general, we can
compute $\PA$ and $\phi_{\rm PA,opp}$ using numerical integrations. To
eliminate the PA rotation from other propagation effects, we simply
use the $\PA$ value from the two poles to calculate Faraday rotation
angle as
\begin{equation}
\phiFR = (\PA-\phi_{\rm PA,opp})/2.
\end{equation}
Figure~\ref{fig:PAvsFreq} shows how the Faraday rotation angle
varies with the wavelength. We find that $\phiFR$ is not
proportional to $\lambda^2$ in a pulsar magnetosphere (see the
left panel of Fig.~\ref{fig:PAvsFreq}). For a set of given pulsar
and plasma parameters, with surface magnetic field
$B_{\ast}=10^{12}$G, the best fit is $\phiFR=0.86\,\lambda^{0.47}$
(see the right panel of Fig.~\ref{fig:PAvsFreq}). The index of
0.47 is different from the theoretical prediction of 1/3 in
eq.~(\ref{eq:dPA3}), probably because of the simplifications of
$F_x$ and $\theta_B$. In real observations, we always calculate
the RM value by
\begin{equation}
{\rm RM}=\frac{\intd\phiFR}{\intd\lambda^2}, \label{eq:rm_obs}
\end{equation}
which varies with $\lambda$ in pulsar magnetosphere as shown in
the middle plane of Figure~\ref{fig:PAvsFreq}. The higher the
observed frequency is, the larger the RM is. Roughly,
RM$\sim0.2\,\lambda^{-1.53}$ can describe RM dependence of
$\lambda$ rather well. For the observation frequency of
$\nu=1.4$GHz (or $\lambda\simeq21$cm), the calculated RM$\simeq 2$
rad m$^{-2}$.

For a higher surface magnetic field, $B_\ast=5\times 10^{12}$G,
the Faraday rotated angles are larger compared with those for
$B_\ast=10^{12}$G (see Fig.~\ref{fig:single_multi}c and
Fig.~\ref{fig:PAvsFreq}). The relation between $\phiFR$ and
$\lambda$ can be fitted to be $\phiFR=2.28\,\lambda^{0.56}$. The
corresponding RM$\sim0.64\,\lambda^{-1.44}$. At $\nu=1.4$GHz,
RM$\simeq 6$ rad m$^{-2}$.

\subsubsection{Phase resolved Faraday Rotation Measure}

\begin{figure}
\centerline{\psfig{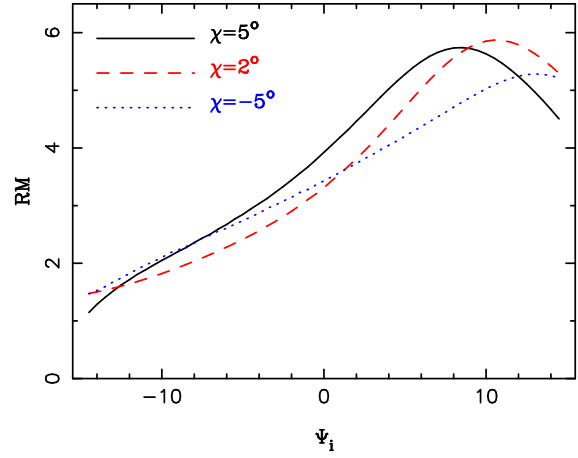}}
\caption{Phase resolved RM from pulsar magnetospheres for
different impact angles of $\chi=5^o$ (solid line), $2^o$ (dashed
line) and $-5^o$ (dotted line).  The pulsar and plasma parameters
for the calculation are the same as used in
Figure~\ref{fig:single}b, except for different initial rotation
phase $\Psi_{\rm i}$ and surface B field
$B_{\ast}=5\times10^{12}$G. } \label{fig:RMvsPsi}
\end{figure}

Having studied the Faraday rotation angles and RMs for photons
propagating through the magnetosphere with given initial pulsar
rotation phases, we can now examine the phase resolved Faraday
rotation effect. As shown in Fig.~\ref{fig:single_multi}f, the
emission height (much less than the light cylinder radius) does not
significantly affect $\phiFR$. Thus, for simplicity, we assume that
all emissions of different rotation phases come from the same height,
at $r_{\rm em}=50R_\ast$. Figure~\ref{fig:RMvsPsi} shows examples of
the Faraday rotation measure as a function of the pulsar rotation
phase. To obtain RM, we integrate the photon polarization evolution
across the magnetosphere, subtract the PA variation caused by other
propagation effects, and then get the Faraday rotated angle $\phiFR$;
we repeat this process with a slightly different frequency, and finally
obtain RM at each rotation phase. From Fig.~\ref{fig:RMvsPsi}, we see
that the leading part of RMs are smaller than the trailing part. The
maximum RM value appears at $\Psi_{\rm i}\simeq8^o$. The RM profiles
change only slightly for different impact angles.

\section{Conclusions}

Motivated by recent observations of phase-dependent Faraday rotation
measures, we have studied the Faraday rotation effect of wave
propagation in pulsar magnetospheres when the natural wave modes are
circularly polarized. Pulsar magnetosphere is filled with relativistic
streaming plasma, the Faraday rotation effect is different from that
of the non-relativistic interstellar medium.  We analyzed the wave
modes and Faraday rotation effect considering the adiabatic evolution
conditions. We used numerical integration of the polarization vector
along the photon ray to incorporate all the propagation effects
self-consistently within a single framework. We find that for highly
symmetric pair plasma, with $|N_p-N_e|\ll N=N_p+N_e$, the
magnetosphere Faraday rotation is negligible. For asymmetric
plasmas (e.g., an electron-ion streams with $N_e\gg N_{\rm GJ}$), the
magnetosphere Faraday rotation effect may be significant. For such an
electron-ion plasma of pulsar magnetosphere, the Faraday
rotation angle $\phiFR$ is not proportional to $\lambda^2$ in pulsar
magnetospheres, but approximately to
$\lambda^{0.5}$.
The induced value of $\RM\propto\lambda^{-1.5}$, becomes larger for higher
frequencies. The phase resolved RMs have smaller values for leading
part and larger values for trailing part. There exists a maximum RM
at the trailing half of the emission beam. The RM profiles do not
change significantly for various impact angles.  The induced rotation
measures for the mainpulse and interpulse have the same magnitude but
opposite sign.

Our calculations in this paper have relied on several simplified
assumptions. For example, we have assumed a cold streaming plasma with
the Lorentz factor $\gamma$ and the density parameter $N/N_{\rm GJ}$
constant throughout the magnetosphere. Note that in reality, even when
the Faraday rotation effect is very weak for a symmetric pair plasma,
other propagation effects can modify the PA curve significantly (see
Wang, Lai \& Han 2010; Andrianov \& Beskin 2010; Beskin \& Philippov
2011).  Overall, our result shows that an unambiguous identification
of the magnetosphere Faraday rotation in the PA data may provide a
valuable probe of the physical condition of pulsar magnetospheres.

\section*{Acknowledgments}

This work has been supported by the National Natural Science
Foundation of China (11003023, 10773016, 10821061 and 10833003).

\label{lastpage}

\end{document}